\documentclass[useAMS,usenatbib]{./mn2e}
\usepackage[dvipdfmx]{graphicx}
\usepackage{epsfig,amsmath,natbib}
\usepackage{color}

\def\be{\begin{equation}} 
\def\ee{\end{equation}} 
\def\ba{\begin{eqnarray}} 
\def\ea{\end{eqnarray}}

\def\kms{\,{\rm {km\, s^{-1}}}} 
\def\cc{\,{\rm {cm^{-3}}}}

\def\HH{${\rm {H_2}}\,\,$}

\def\HI{\hbox{H$\scriptstyle\rm I$}}

\def\CII{\hbox{C~$\scriptstyle\rm II\ $}}

\def\gsim{\lower.5ex\hbox{\gtsima}} 
\def\lsim{\lower.5ex\hbox{\ltsima}} \def\gtsima{$\; \buildrel > \over 
\sim \;$} \def\ltsima{$\; \buildrel < \over \sim \;$} \def\prosima{$\; 
\buildrel \propto \over \sim \;$} \def\gsim{\lower.5ex\hbox{\gtsima}} 
\def\lsim{\lower.5ex\hbox{\ltsima}} 
\def\simgt{\lower.5ex\hbox{\gtsima}} 
\def\simlt{\lower.5ex\hbox{\ltsima}} 
\def\simpr{\lower.5ex\hbox{\prosima}} \def\la{\lsim} \def\ga{\gsim} 
  
 \def\gtsima{$\; \buildrel > \over \sim \;$} 
\def\ltsima{$\; \buildrel < \over \sim \;$} 
\def\gsim{\lower.5ex\hbox{\gtsima}} 
\def\lsim{\lower.5ex\hbox{\ltsima}} 
\def\simgt{\lower.5ex\hbox{\gtsima}} 
\def\simlt{\lower.5ex\hbox{\ltsima}} 
\def\simpr{\lower.5ex\hbox{\prosima}} 
\def\la{\lsim} 
\def\ga{\gsim}

\def\E3{{\cal E}_{\rm g}^{III}}

\def\Msun{M_\odot}

\def\x12{x_{1/2}} 

\def\Tcmb{T_{\rm CMB}}

\title[The infrared-dark dust content of high redshift galaxies] {The infrared-dark dust content of high redshift galaxies} 
\author[Ferrara et al.]{A. Ferrara$^{1,4}$, H. Hirashita$^{2}$, M. Ouchi$^{3,4}$, S. Fujimoto$^3$ \\
$^{1}$ Scuola Normale Superiore, Piazza dei Cavalieri 7, I-56126 Pisa, Italy\\
$^{2}$ Institute of Astronomy and Astrophysics, Academia Sinica, No.\ 1, Sec.\ 4, Roosevelt Rd., Taipei 10617, Taiwan\\
$^{3}$ Institute for Cosmic Ray Research, The University of Tokyo, Kashiwa-no-ha, Kashiwa 277-8582, Japan\\
$^{4}$ Kavli IPMU, WPI, The University of Tokyo, Kashiwa, Chiba 277-8583, Japan\\
}

\begin{document} 
 
\date{\today} 
 
\pagerange{\pageref{firstpage}--\pageref{lastpage}} \pubyear{2012} 
 
\maketitle 
 
\label{firstpage} 
\begin{abstract} 
We present a theoretical model aimed at explaining the IRX-$\beta$ relation for high redshift ($z \simgt 5$) galaxies. Recent observations \citep{Capak15, Bouwens16} have shown that early Lyman Break Galaxies, although characterized by a 
large UV attenuation (e.g. flat UV $\beta$ slopes), show a striking FIR deficit, i.e. they are ``infrared-dark''. This marked deviation 
from the local IRX-$\beta$ relation can be explained by the larger molecular gas content of these systems. While dust in the diffuse ISM attains relatively high temperatures ($T_d \simeq 45$ K for typical size $a=0.1 \mu$m; smaller grains can reach $T_d \simgt 60 $ K), a sizable fraction of the dust mass is embedded in dense gas, and therefore remains cold. If confirmed, the FIR deficit might represent a novel, powerful \textit{indicator of the molecular content of high-$z$ galaxies} which can be used to pre-select candidates for follow-up deep CO observations. Thus, high-$z$  CO line searches with ALMA might be much more promising than currently thought. 
\end{abstract}


\section{Motivation}
\label{Mot}
Dust grains are a fundamental constituent of the interstellar medium (ISM) of galaxies. A large fraction ($\approx $ 50\%  in the Milky Way) of the heavy elements produced by nucleosynthetic processes in stellar interiors can be locked into these solid particles. They are vital elements of the ISM multiphase gas life-cycle and key species for star formation, as they catalyze the formation of molecules on their surfaces. Most relevant here, they efficiently absorb optical/ultraviolet (UV) stellar light, by which they are heated, and re-emit this energy as longer (far-infrared, FIR) wavelength radiation that can freely escape from the galaxy. It is then natural to expect a tight relation between the UV ``deficit'' and the IR excess produced by this process.

Indeed such correlation, called the IRX-$\beta$ relation \citep{Meurer95, Meurer99, Calzetti00, Takeuchi12, Grasha13}, links the UV spectral slope (defined as $\propto \lambda^\beta$) with the FIR excess, i.e. the ratio between the total FIR and $\approx 1500$ \AA\ UV fluxes, $F_{FIR}/F_{UV}$. Such relation has been extensively applied to the local Universe, and more recently extended to redshift $z \approx 2$ \citep{Reddy12, Alvarez16, Fujimoto16}.  The relation has proven to be very robust, at least for starburst galaxies, and useful as it allows to estimate the total UV attenuation. Conversely, if the intrinsic and observed $\beta$ values are known, one can derive the expected total FIR luminosity. 

The presence of dust at high ($z\simgt 6$) redshift implies that conventional dust sources (AGB and evolved stars) are not the dominant contributors. This is because their evolutionary timescales are close or exceed the Hubble time at that epoch ($\approx 1$ Gyr). Following the original proposal by \cite{Todini01}, it is now believed that the first cosmic dust were formed in the supernova ejecta ending the evolution of much more fast-evolving massive stars \citep{Hirashita02, Nozawa07,  Bianchi07, Gall11}. For similar reasons the standard grain growth suffered by grains during their residence time in molecular clouds (MC) of contemporary galaxies cannot increase the amount of dust by considerable amounts \citep{Ferrara16}. Thus, albeit quasar host galaxies show remarkably high dust masses \citep{Beelen06, Michalowski10}, in general the dust-to-gas ratio towards high-redshift rapidly decreases  \citep{Dunlop13} as also witnessed by the observed steepening of early galaxies UV spectra. This does not come as a complete surprise given that the average metallicity of the Universe increases with time. 

\cite{Ferrara99} (for a recent calculation see \cite{DaCunha13}) noticed another important feature of high-$z$ dust. Due to the redshift increase of CMB temperature, $\Tcmb= T_0 (1+z)$ K with $T_0=2.725$, the FIR signal from dust becomes increasingly swamped by the CMB. At $z=6$, for example, $\Tcmb=19.07$ K; as usually dust temperatures in the \textit{diffuse}   
ISM of galaxies are in the range $20-40$ K, the effect cannot be neglected. Even more dramatic, if not complete, might be the suppression of the signal from dust in dense regions (e.g. molecular clouds) where the dust is in thermal equilibrium with the CMB.

In the light of these physical ideas, here we intend to revisit the interpretation of recent dust detections at redshift $z\simgt 5$. The superb sensitivity of the ALMA interferometry has allowed to detect the FIR signal of a handful of Lyman Break Galaxies (LBGs) for which HST rest-frame UV photometry (and hence a $\beta$ determination) is available \citep{Capak15}. This experiment has reported a puzzling deviation of detected LBGs from the more local IRX-$\beta$ relation. In practice, these galaxies, although characterized by relatively flat $\beta \approx -1$ values, indicative of non-negligible dust attenuation, show a noticeable FIR deficit, i.e. they are relatively ``infrared-dark''.   

Such suggested deficit has been strongly reinforced by an even more recent report by the ASPECS survey \citep{Bouwens16}. The authors have performed deep 1.2 mm-continuum observations of the Hubble Ultra Deep Field (HUDF) to probe dust-enshrouded star formation from 330 LBGs spanning the redshift range $z = 2-10$. The striking result is that the expectation from the Meurer IRX-$\beta$ relation at $z=4$ was to detect at least 35 galaxies. Instead, the experiment only provided 6 tentative detections (in the most massive galaxies of the sample). Clearly, redshift evolution either of the dust temperature and/or mass must play a key role. Understanding this behavior is the central aim of the present study\footnote{Throughout the paper,  we assume a flat Universe with the following cosmological parameters:  $\Omega_{\rm m} = 0.308$, $\Omega_{\Lambda} = 1- \Omega_{\rm m} = 0.692$, and $\Omega_{\rm b} = 0.048$,  where $\Omega_{\rm M}$, $\Omega_{\Lambda}$, $\Omega_{\rm b}$ are the total matter, vacuum, and baryonic densities, in units of the critical density,   and  $h$ is the Hubble constant in units of 100 km/s \citep{Ade15}.  }.

An exception to the above scenario is the puzzling case of A1689-zD1 \citep{Watson15, Knudsen16}, a $z=7.5\pm 0.2$ gravitationally-lensed
LBG
for which the thermal dust emission has been detected by ALMA. The large FIR flux $L_\mathrm{FIR}=(6.2\pm 0.8)\times 10^{10}$ L$_{\sun}$ is indicative of considerable amounts of dust, consistent with a Milky Way dust-to-gas ratio. The rest-frame UV slope is estimated to be $\beta=-2\pm 0.1$.    

In this paper we propose a novel idea to explain the physics behind the IRX-$\beta$ relation evolution. Moreover, we will introduce a new indicator which can be used to infer the molecular mass of early galaxies in a regime where classical tracers as CO emission lines might be unavailable, difficult to obtain, or affected by increasing uncertainties (as, e.g., the CO-to-H$_2$ conversion factor).

\section{Dust mass derivation}
\label{Met}

Assume that the\textit{ intrinsic} UV galaxy emission spectrum is $F_\lambda^i \propto \lambda^{\beta^i}$. Such spectrum is modulated by the dust extinction curve via the optical depth $\tau_\lambda $
Then, the emerging (i.e. observed) spectrum is simply given by $F_\lambda^e \propto F_\lambda^i e^{-\tau_\lambda}$. Write the definition of the \textit{observed} UV slope $\beta$ as
\be
\beta = {{\log F_{\lambda_1}^e - \log F_{\lambda_2}^e}\over {\log {\lambda_1} - \log {\lambda_2}}},
\label{betadef}
\ee  
where $(\lambda_1, \lambda_2) = (1600, 2500)$ \AA. From this definition it is straightforward to obtain a relation between $\beta$, $\beta^i$, and the standard V-band optical depth $\tau_V$:
\be
\beta  = \beta^i - \left({\tau_{\lambda_1}\over \tau_{V}}-{\tau_{\lambda_2} \over \tau_{V}} \right) \tau_V {\log e \over \log (\lambda_1/\lambda_2)},
\label{beta}
\ee
where the optical depths are normalized to the value at the $V$ band (5500 \AA).
We adopt the SMC extinction curve as it often used as a representative of a curve with a steep wavelength dependence \citep[e.g.][]{Capak15}; if we adopt a flatter extinction curve, the requirements on the ``infrared-dark'' dust component obtained in this paper should be taken as lower bounds. For the SMC curve given by \cite{Gordon03}, it is $(\tau_{\lambda_1}/\tau_V, \tau_{\lambda_2}/\tau_V)=(4.4, 2.5)$; then  
$\tau_V=0.25(\beta-\beta^i)$.  

We now show that $\beta$ can be simply linked to the dust mass per unit area, $\Sigma_d$, in the galaxy. Recall that $A_V = 1.086 \tau_V$, and then write the following equality:
\be
{{A_V}\over {\Sigma_d}} = {{A_V}\over {N_H}} {{N_H}\over {\Sigma_d}},
\label{AV}
\ee  
where $N_H$ is the total hydrogen column density on the galaxy. Adopt \cite{Weingartner01} grain size distribution which has been shown to reproduce (among others) the SMC extinction curve. This study finds that for the SMC  ${A_V}/ {N_H}=6.2 \times 10^{-23} \mathrm{\rm cm}^2$. They also obtain the total grain volumes, $V_g$ ($V_s$), per H-atom in the carbonaceous (silicate) populations, normalized to their abundance/depletion-limited values $(0.518, 0.745)\times 10^{-27} \cc \mathrm{H}^{-1}$ carbonaceous and silicate grains, respectively. In these units \cite{Weingartner01} find $(V_g, V_s) = (0.254, 1.308)$. Take the material density of carbonaceous and silicate grains to be  $(\delta_g, \delta_s) = (2.24, 3.5) \cc$. By summing the mass of the two species, we obtain $\Sigma_d/N_H = 3.7\times 10^{-27}  \mathrm{ g~H}^{-1}$ with a silicate/carbonaceous mass ratio $\approx 11:1$.

By substituting the previous expressions into eq. (\ref{AV}), we finally obtain
\be
{{A_V}\over {\Sigma_d}} = 3.49\times 10^{-6} \textrm{mag} \left({M_\odot \over \textrm{kpc}^2}\right)^{-1}.
\label{AV1}
\ee  
Combine this expression with the $\beta - \tau_V$ relation to obtain
\be
\Sigma_d = 7.62 \times 10^{4} (\beta - \beta^i) {M_\odot \over \textrm{kpc}^2}.
\label{Sigmad}
\ee  
The intrinsic slope appropriate for $z\simgt 5 -6$ galaxies is $\beta^i = -2.5$ \citep{Bouwens14}, although some uncertainty remains. Such value enters only as an offset (see eq. \ref{beta})  in our curves and does not effect any of the qualitative conclusions drawn here. Thus, we will use this value for numerical estimates from now on. 


\section{Dust temperature derivation}
Energy absorbed by dust from UV radiation is rapidly thermalized. Grains attain an equilibrium temperature, $T_d$, which typically implies re-emission in the FIR. The grain temperature depends on its radius, $a$, absorption/emission coefficients, and UV flux; it can be computed by balancing the rate of energy absorption, $\dot E_+$, with that of emission, $\dot E_-$. The former can be written as
\be
\dot E_+^j =  \int_{912\,\mathrm{\AA}}^{4000\,\mathrm{\AA}} \pi a^2 Q_{abs}^j(\lambda, a) \dot \Sigma_{\lambda} d\lambda ,
\label{Balance}
\ee  
where $Q_{abs}^j(\lambda, a)$ is the dust absorption coefficient (taken from \cite{Draine84}), and $j=c,s$ (i.e., carbonaceous,
silicate). Note that the grain size distribution, $dN/da$ is different for carbonaceous and silicate grains (see below). LyC photons
($\lambda < 912$ \AA) are readily absorbed by \HI; those with $\lambda > 4000$ \AA\ contribute negligibly to grain heating in actively star-forming systems \citep{Buat96}. Finally, $\dot
\Sigma_{\lambda}$ is the specific UV surface luminosity\footnote{We assume that $L_{1600}$ corresponds to the \textit{observed} specific   luminosity. This is slightly inconsistent as the radiation field  inside the galaxy is somewhat higher, being modulated by dust
  absorption itself. Here, we neglect this complication as in most  cases of interest $\tau_V \ll 1$. Moreover, as shown later, the IRX does not depend on the UV field adopted.}, $\dot \Sigma_{\lambda}=L_{1600}
(\lambda/1600\,\mathrm{\AA})^{\beta^i}/\pi r_d^2$. The radius $r_d$ is the ``effective'' extent of dust distribution and we take it as $\propto
r_e$. Both $r_e$ and $r_d$ are obtained from averaging over the galaxies in the Capak sample, further assuming that $r_d$ is traced by
[\CII]158$\mu$m line emission. We define both radii as one half of the geometric mean of the minor and major axes, as measured. For the 8 galaxies for which both the angular sizes of [\CII]158$\mu$m and UV emitting regions have been determined, the average is $r_d/r_e=1.7$.

The grain will radiate energy at a rate given by 
\be
\dot E_+^j   =  \int_0^\infty 4\pi\kappa^j_{\lambda} (\frac{4}{3}\pi a^3 \delta^j) B_\lambda (T_\mathrm{d})\,\mathrm{d}\lambda,
\ee
where $B_\lambda (T_\mathrm{d})$ is the Planck function. Write the FIR absorption coefficient as $\kappa_\lambda =\kappa_{158} (\lambda /\lambda_{158})^{\beta_d}$, where $\kappa_{158} = 1.90\times 10^{12}$ Hz (\citet{Draine84}) is the appropriate value for 
carbonaceous/silicate grains at 158 $\mu$m. The equilibrium temperature, $T_\mathrm{d}(a, \beta, \dot \Sigma_\lambda$), can then be obtained by imposing $\dot E_+=\dot E_-$.

\begin{figure}
\center
\includegraphics[width=80mm]{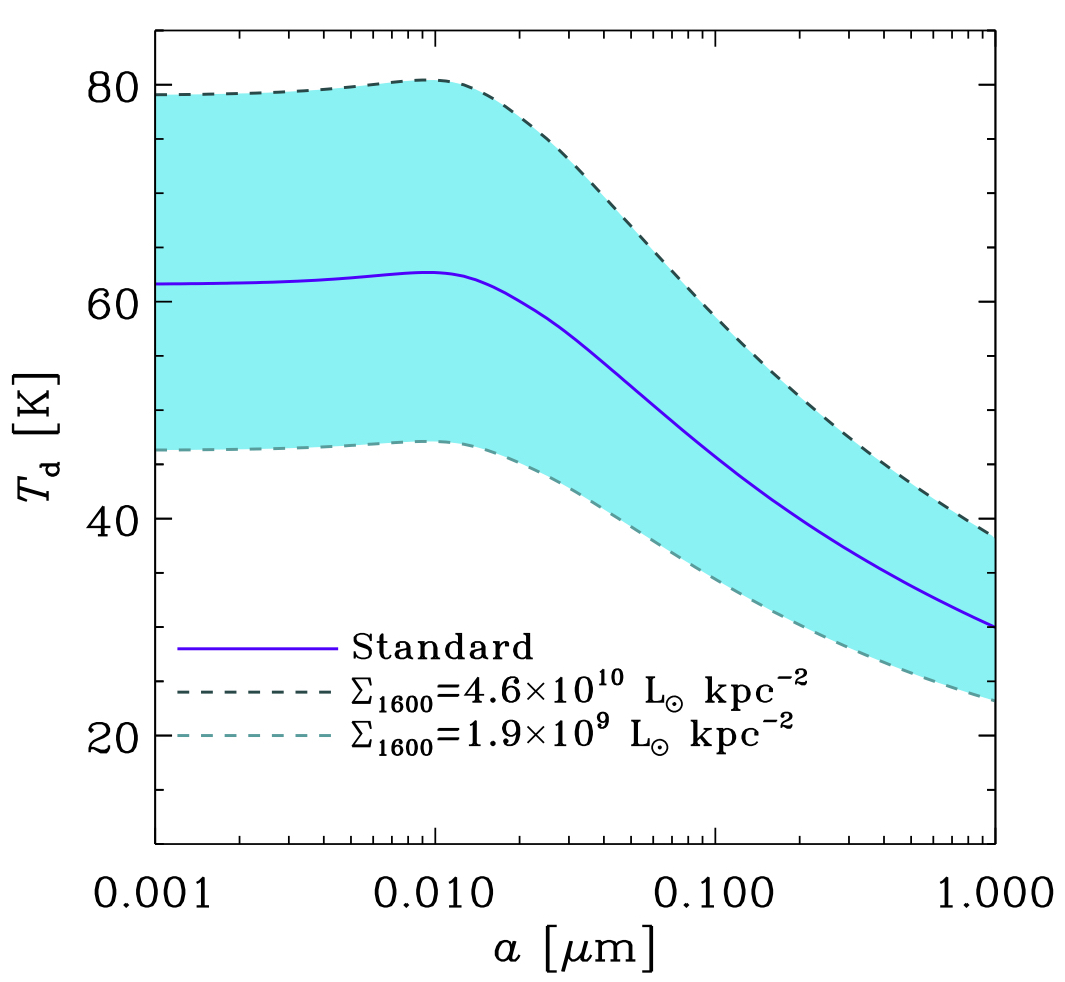}
\caption{Temperature of silicate dust grains residing in the ISM as a function of grain radius. The solid curve is the standard case, corresponding to the mean $\dot \Sigma_{\lambda}$ values measured in the Capak sample.
Dashed lines and the enclosed shaded region represent the full variation range of the UV luminosity (see text for values). }     
\center
\label{Fig1}
\end{figure}

The results are shown in Fig. \ref{Fig1} as a function of grain radius and for range UV surface luminosity ($\log (\lambda \dot \Sigma_{\lambda}/L_\odot {\rm kpc}^{-2}) =10.02^{+0.65}_{-0.75}$ at 1600 \AA) deduced from the Capak galaxy sample\footnote{We have removed from the sample the galaxy HZ5 known to be a X-ray detected quasar}.
For conciseness, we present results for the silicate grains, the dominant species; differences with carbonaceous grains are minor anyway. We define as ``standard'' the curve corresponding to the mean values of the UV surface luminosity. We find that in the standard case typical grains ($a\approx 0.1 \mu$m) have temperatures $T_d \simeq 45$ K, but smaller grains can reach $T_d \simgt 60 $ K. As expected, instead, larger grains are colder.
Temperatures from about $30-60$ K for $a=0.1\, \mu$m are found as a result of a UV field intensity variation of a factor 24, roughly consistent with a temperature dependence $(\dot \Sigma_{\lambda})^{1/6}$. Obviously, large intensities produce hotter dust.

We emphasize, in agreement with the findings by \cite{Bethermin15}, the important point that dust in high-$z$ galaxies is substantially hotter than observed locally. This temperature increase is caused by the larger star formation rate per unit areas characterizing early systems, and correspondingly higher UV surface luminosities.   

\subsection{Dust in molecular clouds}
The previous calculation applies to dust located in the diffuse ISM. Grains embedded in MCs are considerably colder. This is because,
due to the much higher gas density, they can effectively self-shield against UV radiation. For example, dust temperatures in the MW
molecular cloud typically do not exceed 20 K \citep{Stutz10, Lippok16}. At high-$z$, though, the CMB temperature sets a minimum temperature floor which is comparable to or higher than that. Thus. $T_d$ drops from the ISM values calculated above to $\Tcmb$ at sufficiently high optical depths inside the cloud. 

To model $T_d$ as a function of radius, we start by defining the properties of a typical MC. Some insight can be gained from the Larson relation \citep{Larson81} which states that cloud sizes are related to their velocity dispersion as $\sigma \propto L^{0.38}$. When combined with the virial equilibrium hypothesis and another scaling relation between density, $n$, and size, $n\propto L^{-1.1}$, the relation implies that the cloud columns density, $N_H = n L\propto L^{-0.1} \approx$ const. The fiducial value for MW clouds is $N_H=10^{22}$ cm$^{-2}$ \citep{Schneider15}, but $\pm 1$ dex variations around this value have been observed. We model MCs as (critical) Bonnor-Ebert (BE) spheres of mass 
\be
M_{BE}= 1.18\, {c_s^4\over p_0^{1/2}G^{3/2}} 
\label{BE}
\ee
where the ISM external pressure, $p_0=\rho_0 c_s^2$ corresponds to a sound speed $c_s$.  Given the BE profile, the key quantity is the column density, $N_H$, of absorbing material penetrated by the UV radiation impinging on the cloud. In Appendix A we show that $N_H$ measured from the cloud surface ($r=r_0$) to radius $0 \le r \le r_0$ is 
\be
N_H(r) = 10^{22} \left({p_0/k_B\over 10^3 \cc K}\right)^{1/2} (\tan^{-1}a-\tan^{-1}ax) \,\rm cm ^{-2}.
\label{NH}
\ee
where $x=r/r_0$. Note that $N_H$ is purely depending on pressure, and that increasing pressure values decrease the cloud mass at constant temperature (i.e. sound speed).

Having defined MC properties, the temperature calculation is the same as in the ISM case, with the only difference that we need to replace $\dot \Sigma_{\lambda}$ in eq. \ref{Balance} with $\dot \Sigma_{\lambda} e^{-\tau_\lambda}$ to account for the attenuation of the UV field as it propagates into the MC. The optical depth is written as
\be
\tau_\lambda = N_H \sum_j \int_{a_{min}}^{a_{max}} \pi a^2 Q_{abs}^j (\lambda, a) {dn^j\over da}(a) da; 
\ee
the grain size distribution, $dn^j/da$ ($j=c,s$), is normalized to a given dust-to-gas ratio, ${\cal D}$:
\be
\sum_j \int_{a_{min}}^{a_{max}} {4\over 3}\pi a^3 \delta^j  {dn^j\over da}(a) da = \mu m_p  {\cal D}.
\ee
Again, we use the \cite{Weingartner01} size distribution with $a_{min} =10$ \AA, and $a_{max} =1\, \mu$m, as prescribed by the \cite{Weingartner01} model. It is useful to introduce the mass absorption coefficient, 
\be
\kappa_\lambda = {\sum_j \int_{a_{min}}^{a_{max}} \pi a^2 Q_{abs}^j (\lambda, a) {dn^j\over da}(a) da \over \mu m_H  {\cal D}}.
\ee
Then the optical depth is simply written as $\tau_\lambda = \kappa_\lambda \mu m_p {\cal D} N_H = \kappa_\lambda \Sigma_d$; $\mu=1.22$ is the mean molecular weight of the gas. Finally, we take ${\cal D} =0.01$, i.e. similar to the MIlky Way value. This is a reasonable assumption as in the star forming regions of early ($z=6-7$) galaxies the metallicity is $Z \simgt 0.1 Z_\odot$ \citep{Pallottini14}.  In addition, we assume that ${\cal D}$ is the same in the two phases, as suggested by several arguments \citep{Goldsmith97, Dobashi08}. 

We pause for a remark. So far we have adopted the same grain size distribution for dust residing in the ISM and in MCs. This is clearly an oversimplification. In reality, due to coagulation and grain growth by accretion of gas phase metals, the distribution is likely to be more skewed towards larger grains. As a result, the $T_d$ values calculated here should be interpreted as upper limits; grains in MCs are likely to be somewhat colder, and their IR emission even more depressed. However, a full treatment of these complex and uncertain processes is outside the scope of the present work and is left to future study.

As a final step, we need to correct the previously computed temperatures to include CMB heating.
The effect of CMB is to provide an extra heating term which does not allow grains to cool below the CMB temperature. 
The corrected temperature is then \citep{DaCunha13}
\be
T_d' = \big\{ T_d^{4+\beta_d} + T_0^{4+\beta_d} [(1+z)^{4+\beta_d}-1]\big\}^{1/(4+\beta_d)}, 
\ee  
where $T_d'$ is the actual temperature of dust including CMB effects.
Clearly, such effects are sub-dominant (but not negligible) as far as ISM dust is concerned (unless star formation rates and UV luminosities are very small), but they can become dramatic in MCs where a large fraction of the dust grains reaches thermal equilibrium with the CMB, causing a drop of its emission, as discussed in the following.

\begin{figure}
\center
\includegraphics[width=80mm]{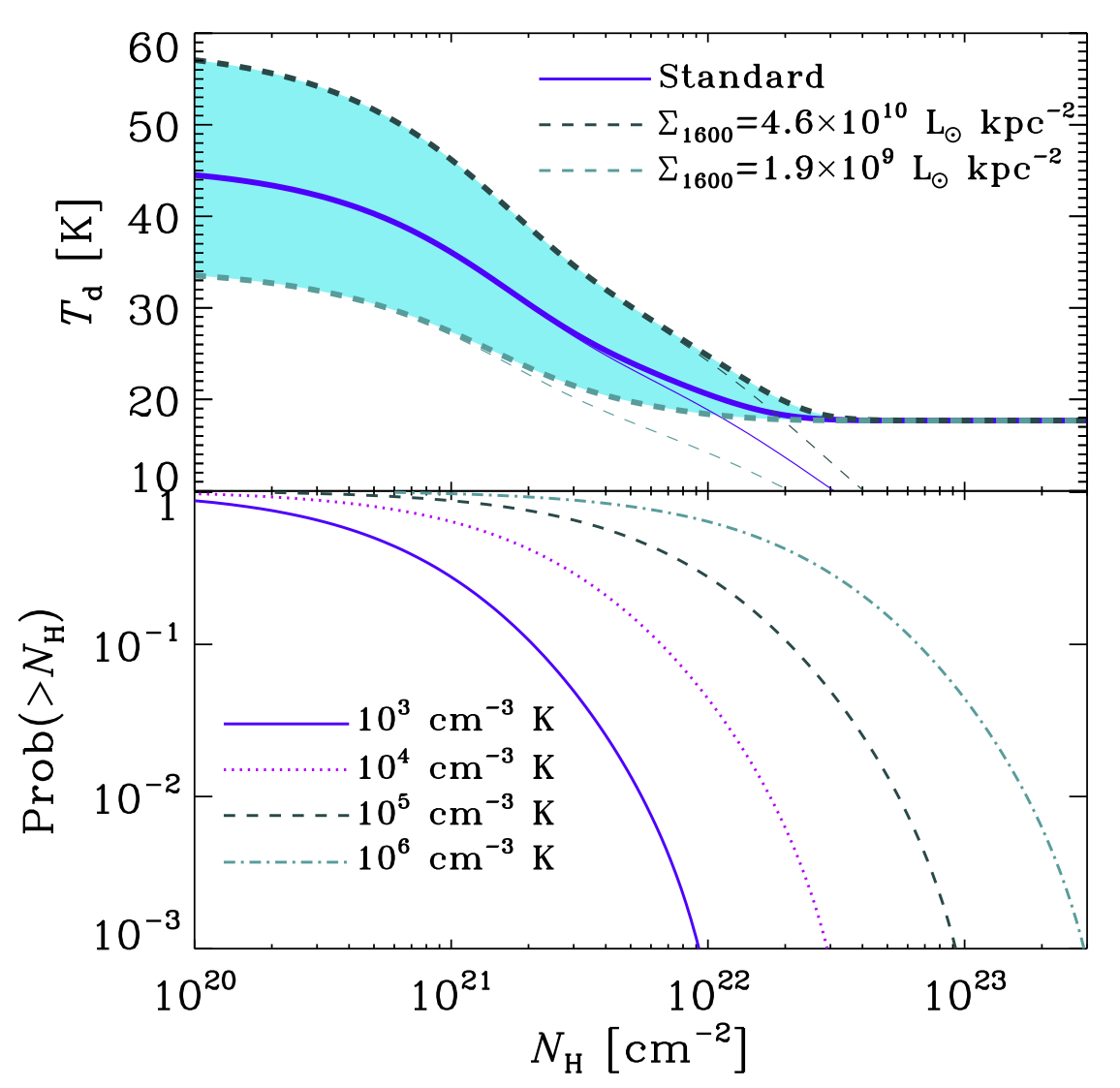}
\caption{{\it Upper panel:} Temperature of $a=0.1\, \mu$m silicate dust grains residing in MC as a function of depth inside the cloud measured by the hydrogen column density, $N_H$.  The solid curves represent the standard case, while the
dashed lines and the enclosed shaded region represent the full variation range of the UV luminosity as shown. The thick and thin lines show the dust temperatures including and excluding CMB heating, respectively. {\it Lower:} Fraction of the total cloud mass located at a depth larger than a given $N_H$ value depending on the ISM pressure values given in the inset.}     
\center
\label{Fig2}
\end{figure}

The results are shown in Fig. \ref{Fig2}. Note that in this case, $T_d$ depends not only on $a$ and $\dot \Sigma_\lambda$, but also on the grain location, measured by the gas column density, within the cloud. Grains located closer to (far from) the cloud surface attain higher (lower) temperatures. For simplicity, we show in Fig. \ref{Fig2} the results for a silicate grain of the typical radius $a=0.1\, \mu$m. As in Fig. \ref{Fig1} we explore the full range of $\dot \Sigma_{\lambda}$ values. Without including the CMB correction, the 
grain temperature drops from the ISM values find before to very low values, $\simlt 10$ K, for $N_H \simgt 3-4\times 10^{22} \mathrm{cm}^{-2}$ for any value of the UV field. However, once CMB (we compute it at $z=5.48$, the mean redshift of the Capak galaxy sample) is included, grains cannot become colder than the CMB photon temperature. This is seen as a flattening of the curves beyond  $N_H = 10^{22} \mathrm{cm}^{-2}$. Note that the CMB effect becomes important at progressively lower $N_H$ as redshift  increases. For example at $z=7$, dust shielded by column densities as low as $N_H \simgt 3\times 10^{21} \mathrm{cm}^{-2}$ sets to the CMB temperature floor.  Temperature differences between carbonaceous and silicate grains are negligible. Considering the 11:1 mass predominance of silicates, we will not differentiate between the two as far as the IR emission is concerned. 
\subsection{Infrared emission by dust}
Having computed the temperature of dust grains both in the ISM and in MC interiors as a function of grain size, absorbing gas column density, and galaxy surface star formation rate we can now predict the integrated infrared emission entering the IRX-$\beta$ relation. 
The total dust emission is the sum of the ISM and MC components, weighted by the molecular gas mass fraction, $\mu=M_{\rm H2}/M_g$, where $M_g$ is the total (i.e. \HH + HI), gas content of the galaxy.

The intrinsic rest-frame luminosity per unit dust mass (erg s$^{-1}$ g$^{-1}$) in a galaxy at redshift $z$ is
\be
{\cal L}^i_\lambda = 4\pi \left[ \mu \kappa_\lambda \langle B_\lambda(T_d')\rangle_{MC} + (1-\mu) \kappa_\lambda \langle B_\lambda(T_d')\rangle_{ISM}\right] ,
\ee   
where we have implicitly assumed that the sub-mm optical depth is $\ll 1$.  The emission for MC is weighted on the dust mass as a function of temperature (see Fig. \ref{Fig2}) and integrated over the grain size distribution. For the ISM it is simply integrated over $dn/da$.
We integrate ${\cal L}^i_{\lambda}$
from 8 to 1000 $\mu$m rest-frame to get the total FIR luminosity per unit mass, ${\cal L}^i$.   

\begin{figure}
\center
\includegraphics[width=80mm]{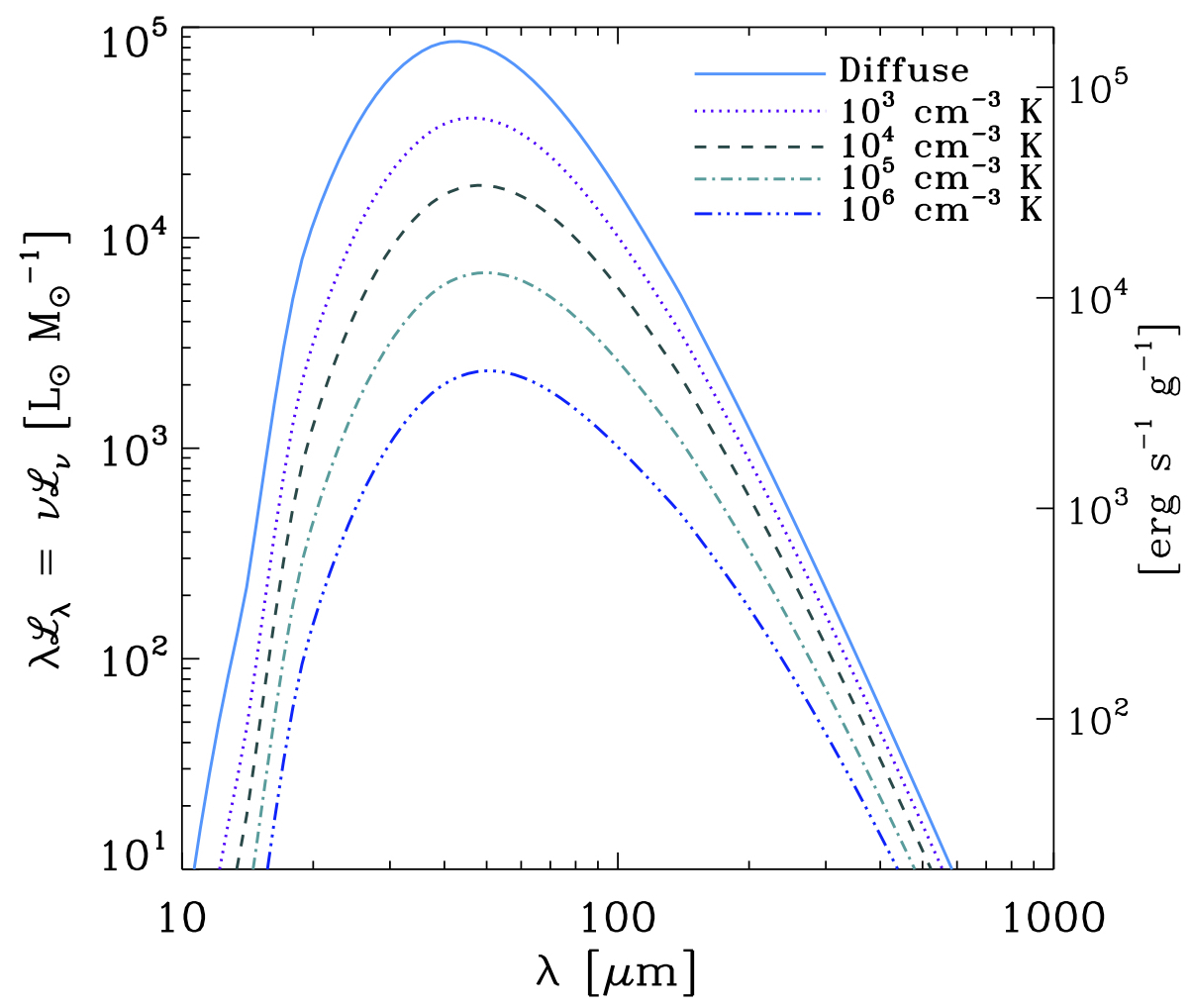}
\caption{Restframe far-infrared Spectral Energy Distribution per dust mass of a typical LBG for different values of the ISM pressure (affecting the emission of molecular clouds). The ``diffuse'' ISM case (solid line), i.e. no molecular clouds, is independent of pressure. The dotted, dashed, dot-dashed, and dot-dot-dot-dashed lines show cases with different ISM pressure, $10^3$, $10^4$, $10^5$, and $10^6$ cm$^{-3}$ K, respectively, for the molecular cloud component ($\mu= 1$). }     
\center
\label{Fig3}
\end{figure}

From this procedure we compute the FIR Spectral Energy Distribution (SED) shown in Fig. \ref{Fig3}. In the case in which the ISM is only made by the diffuse components, we expect a peak luminosity per unit dust mass of about $10^5 L_\odot$. Such maximum is located at $\lambda \approx 40 \mu$m, reflecting the moderately high dust temperatures that we found above. As a fraction of the ISM dust is stored into MCs, the FIR emission depends on ISM pressure, which regulates MC properties (see eq. \ref{NH}). As pressure is increased (Fig. \ref{Fig3}, curves for $\mu=1$), $N_H$ becomes larger, shielding a progressively higher fraction of the dust mass from UV radiation and heating. As a consequence, the SED drops by three orders of magnitude for $p_0/k_B=10^6 \cc$K, and slightly shifts to longer wavelengths.

In addition to modifying the dust temperature, the CMB introduces a second important effect. As the dust emission is observed against the CMB, the intensity of the latter must be subtracted out. 
Assuming that the observed wavelength corresponds to $\lambda_0$ (i.e. the observed wavelength is related the intrinsic one$, \lambda$, by $\lambda_0=\lambda(1+z)$), the subtraction of the CMB diminishes fraction
$B_\lambda(\Tcmb)/({\cal L}_\lambda^i/4\pi\kappa )$ of the dust emission \citep{DaCunha13}.
Then the observable luminosity after the subtraction of the CMB at wavelength $\lambda_0$ is
\be
{\cal L}^o=\left[  1 - {B_\lambda(\Tcmb)\over B_\lambda(T_d')} \right] {\cal L}^i .
\ee

\section{The IRX-$\beta$ relation}
\label{Res}
\begin{figure}
\center
\includegraphics[width=80mm]{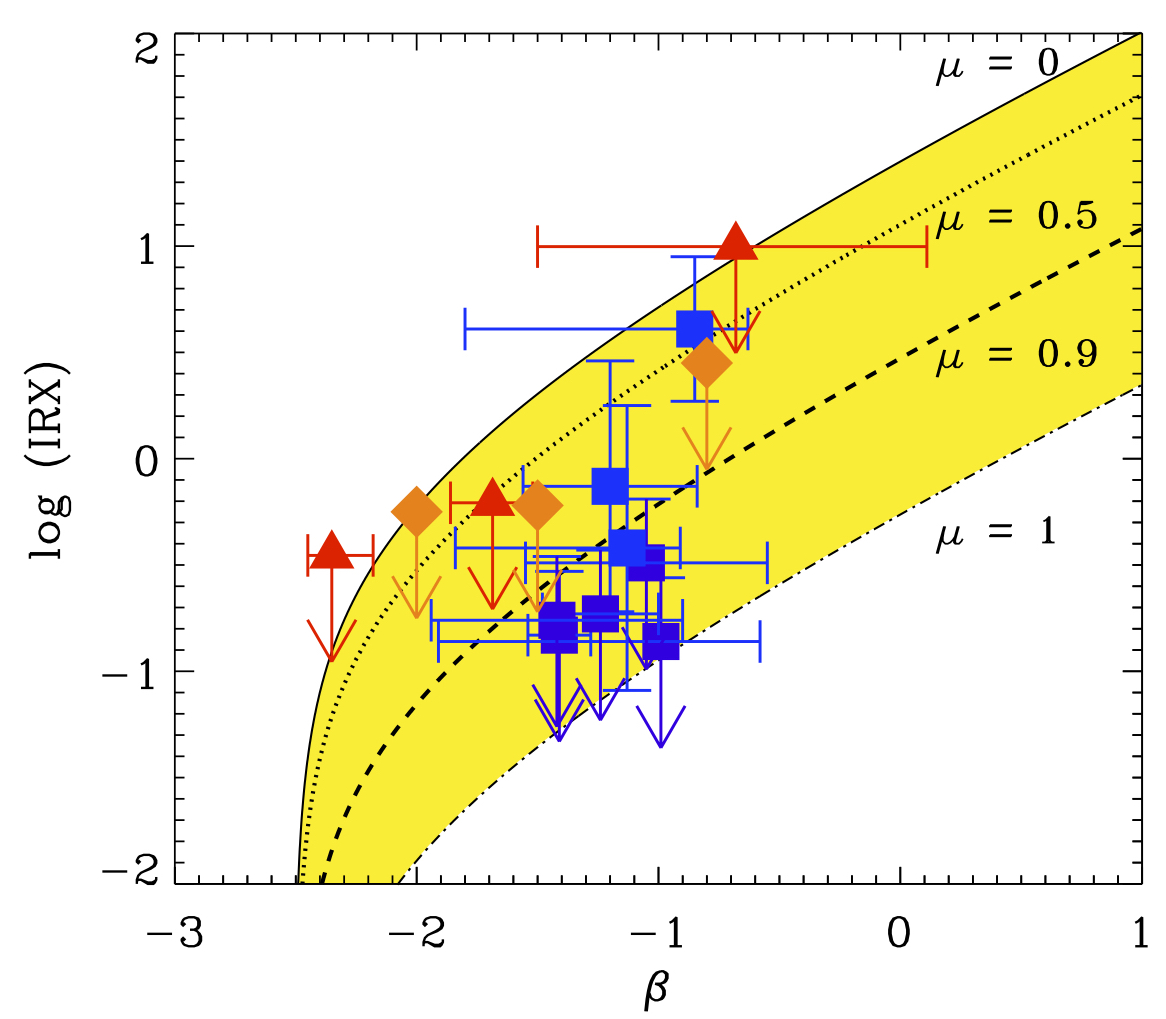}
\caption{IRX-$\beta$ relation (assuming a SMC extinction curve) corresponding to an ISM pressure $p_0/k_B=10^6\, \mathrm{K} \cc$  and dust-to-gas ratio ${\cal D} =0.01$. Curves refer to different values of the molecular gas fraction, $\mu=0, 0.5, 0.9$, and 1 for the solid, dotted, dashed, and dot-dashed lines, respectively. The case $\mu=0$ corresponds to diffuse ISM only, i.e. no molecular component, while $\mu =1$ corresponds to the fully molecular case. Points represent data from \cite{Capak15} (squares) for LBGs at $z=5$--6, \cite{Bouwens16} (diamonds) at $z=4$--10 in the Hubble Ultra Deep Field (HUDF) (2 $\sigma$ upper limits after stacking), 
and Fujimoto et al., in prep. (triangles) at $z=5$--9 in the HUDF (2 $\sigma$ upper limits  after stacking). Data points assume $T_d = 35$ K.}     
\center
\label{Fig4}
\end{figure}

The final step is to combine the FIR emission calculated above with the $\beta$ value to get the IRX-$\beta$ relation. 
In practice, this entails computing for a given $\beta$ the dust surface density via equation (\ref{Sigmad}) to get the total FIR surface luminosity $\Sigma_d{\cal L}^o$. This is subsequently normalized to the specific UV surface luminosity, $\dot \Sigma_\lambda\exp(-\tau_{1600})$. The curves in the IRX-$\beta$ plane are independent on $\dot \Sigma_\lambda$. This is because the dust IR emission is calculated consistently with the incident UV field. The results are shown in Fig. \ref{Fig4} for a pressure value $p_0/k_B=10^6 \cc$K consistent with those found in the central regions of simulated $z=6$ galaxies (\cite{Gallerani16}; Pallottini et al. in prep.).

We start by noting that our model for the purely diffuse ISM case (solid line in Fig.\ \ref{Fig4}) reproduces very well the empirically-derived IRX-$\beta$ curve for the SMC extinction curve (consistent with our assumption) shown in Capak et al.\ (2015). 
This is not a trivial result in itself, as it lends theoretical support to such observational relation. Note that the low-$\beta$ drop-off of the curve is set by the assumed intrinsic spectral slope, which we have assumed\footnote{We warn that the uncertainty in $\beta_i$ makes it difficult to put firm constraints on the precise value of $\mu$. However, $\mu$ affects the IRX--$\beta$ relation also for $\beta\approx \beta^i$, indicating that MC effects cannot be neglected even for low-extinction objects.} to be $\beta^i=-2.5$.

The diffuse ($\mu=0$) curve sets essentially an upper bound to the IRX value. However, at a fixed $\beta > \beta_i$, lower IRX values can be obtained (and indeed observed, see Capak et al. 2015, Bouwens et al. 2016, Fujimoto et al. in prep.; see data points in Fig.  \ref{Fig4} ). These correspond to galaxies in which a certain fraction $0 < \mu \le 1$ of their gas is in dense, molecular form. For example, these selected values of the molecular fraction $\mu=(0, 0.5, 0.9, 1)$ yield, for $\beta=-1$, $\log({\rm IRX})= (0.65, 0.3,-0.3, -1.0)$, corresponding to a total variation of a factor 47. 
Clearly, as the lowest IRX value (IRX=0.1) corresponds to the limiting case in which all the dust is located in shielded molecular clouds, it must be interpreted as a lower bound. The IRX variation interval due to the ``hidden dust'' effect we are advocating become even larger for smaller $\beta$, and it spans about a factor 100 at $\beta=-2$. Note that all the measured data points and upper limits nicely fall in between the $0 < \mu < 1$ curves. This is a successful consistency test of the model. 

Hence, the molecular content of the galaxy has a strong impact on the dust continuum emission for any value of $\beta$, and noticeably also for galaxies whose dust content is not very large (small $\beta$). It follows that it could become increasingly important at high-$z$. The interesting feature is that the suppression of the FIR emission from such systems has already been reported in the literature, rising thorny questions about the standard interpretation in which molecular clouds are not considered. ISM pressure plays a key role; while CMB introduces variations $\le 50\%$ in the IRX relation, its effects is subdominant with respect to the shielding suffered by dust in MCs. We conclude that the present model offers a simple and physical explanation for the experimental evidence.  

As one of the major finding of this work is that the dust temperature in high-$z$ galaxies is higher than locally, an alternative scenario for the explanation of the FIR deficit can also be considered, as pointed out by \cite{Ouchi99, Bouwens16}. Essentially, a biased-low dust temperature assumption would imply a lower luminosity, and therefore a lower IRX value. On the contrary, higher luminosities would arise from hotter dust. According to this scenario, in case of hotter dust the observational data points could be shifted upward by some amount and become more consistent with the SMC relation. Indeed, this is a possibility envisaged by our model for the $\mu=0$ case (no molecular gas, all dust is hot and in the diffuse ISM phase). If dense clouds exist, though, they would decrease the mass-weighted dust temperature, essentially justifying the current flux-luminosity conversion using $T_d = 35$ K. These two scenarios have distinct predictions (see next paragraph) that can be tested with, e.g. sub-mm observations.  

A very relevant outcome of the present model is the huge sensitivity of the IRX-$\beta$ relation to $\mu$. As galaxies with a larger molecular content should be characterized by a lower IRX, our model suggests an interesting way to pre-select candidates for molecular studies at high-$z$ studies. So far, no normal (LBG or LAE) galaxy has been observed at $z>5$. Such an experiment is very challenging, and the only two LAEs (HCM 6A, IOK-1) at $z\approx 6.5$ for which deep CO(1-0) observations have been attempted \citep{Wagg09} have provided only upper limits. However, with the capabilities of ALMA such experiments are now becoming not only possible, but also very urgent to complement other FIR emission lines recently observed, e.g. \cite{Maiolino15,Inoue16}. Enabling the study of the molecular component of these early galaxies via, e.g. high-$J$ lines of the CO molecule, is a fundamental step to understand many of their peculiar properties. By selecting $\beta$ and IRX values maximizing the molecular fraction $\mu$ shown in Fig. \ref{Fig4}, one can efficiently and reliably pre-select suitable candidates for follow-up ALMA observations.     
%


\section{Conclusions}
\label{Con}
We have presented a dust extinction and FIR emission model that is aimed at explaining the IRX-$\beta$ relation for high redshift ($z \simgt 5$) galaxies. We have first derived the dust mass vs. UV spectral slope, $\beta$ relation. Then the temperature of grains exposed to the typical range of internal UV field intensities of $z\simeq 6$ LBG galaxies has been computed. Such calculation allows for attenuation of the UV flux inside molecular clouds and account for the CMB effects both for what concerns dust temperature and dust continuum suppression. When combined, these two results allow to predict the expected IRX-$\beta$ relation (assuming a SMC extinction curve) and compare it with recent data. 

A key result is that our model reproduces extremely well (essentially without free parameters) the striking FIR deficit observed, showing that these early systems are ``infrared-dark''. We suggest that the deficit is caused by an increasing  molecular gas content of these systems. While dust residing in the diffuse ISM attains large temperatures  ($T_d \simeq 45$ K for typical size $a=0.1 \mu$m; smaller grains can reach $T_d \simgt 60 $ K) due to presence of intense interstellar  UV fields
\citep{Bethermin15}, dust located in molecular clouds becomes very cold (but not colder than the CMB). As galaxies with a larger molecular content should be characterized by a lower IRX, our model suggests an interesting way to pre-select candidates for molecular studies at high-$z$ studies. In other words, for a given value of $\beta$, we predict that the galaxies with the largest molecular content are those characterized by the lower IRX values. That is, IRX anti-correlates with the molecular fraction, $\mu$.
Thus, searching for CO line emission (and perhaps H$_2$ lines directly with, e.g., SPICA\footnote{www.ir.isas.jaxa.jp/SPICA/}) from high-$z$ galaxies with ALMA might be much more promising than currently thought. 

A first attempt along these lines has been presented by \cite{Riechers14} who observed the LBG galaxy HZ6 (IRX = 0.38, $\beta=-1.13$), also part of the Capak sample. The authors observed HZ6 with the VLA in search for CO($J = 2-1$) line emission, obtaining an approximate 3$\sigma$ limit of $<0.03$ Jy km s$^{-1}$. However, higher-$J$ transitions, like the CO(6-5), detectable with ALMA, are expected to be considerably more luminous \citep{Vallini16}.  

If successful these experiments could provide a fundamental insight on the highly debated processes of dust formation and growth \citep{Ferrara16}, on the evolution of the molecular content of galaxies \citep{Genzel15} and the relationship between the two.

\section*{Acknowledgments} 
We thank R. Bouwens and the ASPECS collaboration for providing data in digital form and S. Gallerani, D. Riechers, S. Salvadori, and F. Walter for useful discussions.  This research was supported in part by the National Science Foundation under Grant No. NSF PHY11-25915. HH is supported by the Ministry of Science and Technology (MoST) grant 102-2119-M-001-006-MY3. This work is supported by World Premier International Research Center Initiative (WPI Initiative), MEXT, Japan, and KAKENHI (15H02064) Grant-in-Aid for Scientific Research (A) through Japan Society for the Promotion of Science (JSPS).


\section*{Appendix A}
\label{App}
We derive here the molecular cloud density profile. We assume that this follow the Bonnor-Ebert solution for an isothermal hydrostatic cloud with external pressure, $p_0=\rho_0 c_s^2$ (see eq. \ref{BE}), which can be well approximated by a flat central core (of radius $r_c$ and density $\rho_c$) + a $r^{-2}$ density profile:
\be
\rho(r) = {2\tilde{\lambda}\rho_c\over 1+ (r/r_c)^2};
\ee
we also allow for a numerical constant $\tilde{\lambda}=3.63$ to normalize the cloud column density to the observed mean value $N_H=10^{22}$ cm$^{-2}$ \citep{Schneider15}. The stability condition of a BE sphere requires that the central density is $\zeta = \rho_c/\rho_0$ times that at the outer edge, $r_0$, with $\zeta=14.1$. This fixes the outer/core radius ratio to $r_0/r_c=\sqrt{2\zeta-1}=\alpha$. Impose that the mass of the cloud, $M_c$, equals the Bonnor-Ebert one (see eq. \ref{BE}), 
\be
M_{BE}\approx 0.9\times10^4 \left({c_s \over 1 \kms}\right) \left({p_0/k_B \over 10^3 \cc K }\right)^{-1/2} \Msun,
\ee
at the outer cloud radius, $r_0$:
\be
4\pi \int_0^{r_0} \rho(r) r^2 dr = 1.18 {c_s^3 \over \rho_0^{1/2} G^{3/2}}.
\label{BE_App}
\ee
This equation can be solved to yield $r_0$ (and therefore the core radius from the previously derived ratio). The outer radius is only function of the external properties of the ISM, and it can be written as
\ba
r_0 &=&  \left[ {1.18\over 8 \pi \zeta\lambda} {\alpha^{3}\over (\alpha -\arctan \alpha)}\right]^{1/3}  {c_s \over \sqrt{G\rho_0} } = 0.497 {c_s \over \sqrt{G\rho_0}}\nonumber\\
     &\approx & 10 \left({c_s \over 1 \kms}\right) \left({n_0 \over 10 \cc }\right)^{-1/2} \rm pc.
\label{BE_App1}
\ea
The above fiducial values of $c_s$ and $n_0$ appropriate for the Cold Neutral Medium (CNM) of the MW give a consistent pressure $p/k_B = 1159 \cc$ K and $M_{BE}=0.8\times 10^4 \Msun$.

It is important to note that to compute the dust temperature in the molecular clouds, the key quantity is the column density of absorbing material penetrated by the UV radiation impinging on the cloud. Following the previous results, the gas column density measured from the cloud surface ($r=r_0$) to radius $0 \le r \le r_0$ is
\be
N_H(r) = \int_{r_0}^r {\rho(r)\over \mu m_p} dr.  
\ee
Thus, the column density is purely a function of pressure. Numerically, this corresponds to
\be
N_H(r) = 10^{22} \left({p_0/k_B\over 10^3 \cc K}\right)^{1/2} (\tan^{-1}\alpha-\tan^{-1}\alpha x) \,\rm cm ^{-2}.
\ee
with $x=r/r_0$.

\bibliographystyle{./apj}
\bibliography{./ref}

\begin{thebibliography}{45}
\expandafter\ifx\csname natexlab\endcsname\relax\def\natexlab#1{#1}\fi

\bibitem[{{{\'A}lvarez-M{\'a}rquez} {et~al.}(2016){{\'A}lvarez-M{\'a}rquez},
  {Burgarella}, {Heinis}, {Buat}, {Lo Faro}, {B{\'e}thermin},
  {L{\'o}pez-Fort{\'{\i}}n}, {Cooray}, {Farrah}, {Hurley}, {Ibar}, {Ilbert},
  {Koekemoer}, {Lemaux}, {P{\'e}rez-Fournon}, {Rodighiero}, {Salvato}, {Scott},
  {Taniguchi}, {Vieira}, \& {Wang}}]{Alvarez16}
{{\'A}lvarez-M{\'a}rquez}, J., {et~al.} 2016, \aap, 587, A122

\bibitem[{{Beelen} {et~al.}(2006){Beelen}, {Cox}, {Benford}, {Dowell},
  {Kov{\'a}cs}, {Bertoldi}, {Omont}, \& {Carilli}}]{Beelen06}
{Beelen}, A., {Cox}, P., {Benford}, D.~J., {Dowell}, C.~D., {Kov{\'a}cs}, A.,
  {Bertoldi}, F., {Omont}, A., \& {Carilli}, C.~L. 2006, \apj, 642, 694

\bibitem[{{B{\'e}thermin} {et~al.}(2015){B{\'e}thermin}, {Daddi}, {Magdis},
  {Lagos}, {Sargent}, {Albrecht}, {Aussel}, {Bertoldi}, {Buat}, {Galametz},
  {Heinis}, {Ilbert}, {Karim}, {Koekemoer}, {Lacey}, {Le Floc'h}, {Navarrete},
  {Pannella}, {Schreiber}, {Smolcic}, {Symeonidis}, \& {Viero}}]{Bethermin15}
{B{\'e}thermin}, M., {et~al.} 2015, \aap, 573, A113

\bibitem[{{Bianchi} \& {Schneider}(2007)}]{Bianchi07}
{Bianchi}, S., \& {Schneider}, R. 2007, \mnras, 378, 973

\bibitem[{{Bouwens} {et~al.}(2016){Bouwens}, {Aravena}, {Decarli}, {Walter},
  {da Cunha}, {Labbe}, {Bauer}, {Bertoldi}, {Carilli}, {Chapman}, {Daddi},
  {Hodge}, {Ivison}, {Karim}, {Le Fevre}, {Magnelli}, {Ota}, {Riechers},
  {Smail}, {van der Werf}, {Weiss}, {Cox}, {Elbaz}, {Gonzalez-Lopez},
  {Infante}, {Oesch}, {Wagg}, \& {Wilkins}}]{Bouwens16}
{Bouwens}, R., {et~al.} 2016, ArXiv e-prints

\bibitem[{{Bouwens} {et~al.}(2014){Bouwens}, {Illingworth}, {Oesch},
  {Labb{\'e}}, {van Dokkum}, {Trenti}, {Franx}, {Smit}, {Gonzalez}, \&
  {Magee}}]{Bouwens14}
{Bouwens}, R.~J., {et~al.} 2014, \apj, 793, 115

\bibitem[{{Buat} \& {Xu}(1996)}]{Buat96}
{Buat}, V., \& {Xu}, C. 1996, \aap, 306, 61

\bibitem[{{Calzetti}(2000)}]{Calzetti00}
{Calzetti}, D. 2000, in Building Galaxies; from the Primordial Universe to the
  Present, ed. F.~{Hammer}, T.~X. {Thuan}, V.~{Cayatte}, B.~{Guiderdoni}, \&
  J.~T. {Thanh Van}, 233

\bibitem[{{Capak} {et~al.}(2015){Capak}, {Carilli}, {Jones}, {Casey},
  {Riechers}, {Sheth}, {Carollo}, {Ilbert}, {Karim}, {Lefevre}, {Lilly},
  {Scoville}, {Smolcic}, \& {Yan}}]{Capak15}
{Capak}, P.~L., {et~al.} 2015, \nat, 522, 455

\bibitem[{{da Cunha} {et~al.}(2013){da Cunha}, {Groves}, {Walter}, {Decarli},
  {Weiss}, {Bertoldi}, {Carilli}, {Daddi}, {Elbaz}, {Ivison}, {Maiolino},
  {Riechers}, {Rix}, {Sargent}, \& {Smail}}]{DaCunha13}
{da Cunha}, E., {et~al.} 2013, \apj, 766, 13

\bibitem[{{Dobashi} {et~al.}(2008){Dobashi}, {Bernard}, {Hughes}, {Paradis},
  {Reach}, \& {Kawamura}}]{Dobashi08}
{Dobashi}, K., {Bernard}, J.-P., {Hughes}, A., {Paradis}, D., {Reach}, W.~T.,
  \& {Kawamura}, A. 2008, \aap, 484, 205

\bibitem[{{Draine} \& {Lee}(1984)}]{Draine84}
{Draine}, B.~T., \& {Lee}, H.~M. 1984, \apj, 285, 89

\bibitem[{{Dunlop}(2013)}]{Dunlop13}
{Dunlop}, J.~S. 2013, in Astrophysics and Space Science Library, Vol. 396, The
  First Galaxies, ed. T.~{Wiklind}, B.~{Mobasher}, \& V.~{Bromm}, 223

\bibitem[{{Ferrara} {et~al.}(1999){Ferrara}, {Nath}, {Sethi}, \&
  {Shchekinov}}]{Ferrara99}
{Ferrara}, A., {Nath}, B., {Sethi}, S.~K., \& {Shchekinov}, Y. 1999, \mnras,
  303, 301

\bibitem[{{Ferrara} {et~al.}(2016){Ferrara}, {Viti}, \&
  {Ceccarelli}}]{Ferrara16}
{Ferrara}, A., {Viti}, S., \& {Ceccarelli}, C. 2016, ArXiv e-prints

\bibitem[{{Fujimoto} {et~al.}(2016){Fujimoto}, {Ouchi}, {Ono}, {Shibuya},
  {Ishigaki}, {Nagai}, \& {Momose}}]{Fujimoto16}
{Fujimoto}, S., {Ouchi}, M., {Ono}, Y., {Shibuya}, T., {Ishigaki}, M., {Nagai},
  H., \& {Momose}, R. 2016, \apjs, 222, 1

\bibitem[{{Gall} {et~al.}(2011){Gall}, {Hjorth}, \& {Andersen}}]{Gall11}
{Gall}, C., {Hjorth}, J., \& {Andersen}, A.~C. 2011, \aapr, 19, 43

\bibitem[{{Gallerani} {et~al.}(2016){Gallerani}, {Pallottini}, {Feruglio},
  {Ferrara}, {Maiolino}, {Vallini}, \& {Riechers}}]{Gallerani16}
{Gallerani}, S., {Pallottini}, A., {Feruglio}, C., {Ferrara}, A., {Maiolino},
  R., {Vallini}, L., \& {Riechers}, D.~A. 2016, ArXiv e-prints

\bibitem[{{Genzel} {et~al.}(2015){Genzel}, {Tacconi}, {Lutz}, {Saintonge},
  {Berta}, {Magnelli}, {Combes}, {Garc{\'{\i}}a-Burillo}, {Neri}, {Bolatto},
  {Contini}, {Lilly}, {Boissier}, {Boone}, {Bouch{\'e}}, {Bournaud}, {Burkert},
  {Carollo}, {Colina}, {Cooper}, {Cox}, {Feruglio}, {F{\"o}rster Schreiber},
  {Freundlich}, {Gracia-Carpio}, {Juneau}, {Kovac}, {Lippa}, {Naab}, {Salome},
  {Renzini}, {Sternberg}, {Walter}, {Weiner}, {Weiss}, \& {Wuyts}}]{Genzel15}
{Genzel}, R., {et~al.} 2015, \apj, 800, 20

\bibitem[{{Goldsmith} {et~al.}(1997){Goldsmith}, {Bergin}, \&
  {Lis}}]{Goldsmith97}
{Goldsmith}, P.~F., {Bergin}, E.~A., \& {Lis}, D.~C. 1997, \apj, 491, 615

\bibitem[{{Gordon} {et~al.}(2003){Gordon}, {Clayton}, {Misselt}, {Landolt}, \&
  {Wolff}}]{Gordon03}
{Gordon}, K.~D., {Clayton}, G.~C., {Misselt}, K.~A., {Landolt}, A.~U., \&
  {Wolff}, M.~J. 2003, \apj, 594, 279

\bibitem[{{Grasha} {et~al.}(2013){Grasha}, {Calzetti}, {Andrews}, {Lee}, \&
  {Dale}}]{Grasha13}
{Grasha}, K., {Calzetti}, D., {Andrews}, J.~E., {Lee}, J.~C., \& {Dale}, D.~A.
  2013, \apj, 773, 174

\bibitem[{{Hirashita} \& {Ferrara}(2002)}]{Hirashita02}
{Hirashita}, H., \& {Ferrara}, A. 2002, \mnras, 337, 921

\bibitem[{{Inoue} {et~al.}(2016){Inoue}, {Tamura}, {Matsuo}, {Mawatari},
  {Shimizu}, {Shibuya}, {Ota}, {Yoshida}, {Zackrisson}, {Kashikawa}, {Kohno},
  {Umehata}, {Hatsukade}, {Iye}, {Matsuda}, {Okamoto}, \&
  {Yamaguchi}}]{Inoue16}
{Inoue}, A.~K., {et~al.} 2016, ArXiv e-prints

\bibitem[{{Knudsen} {et~al.}(2016){Knudsen}, {Watson}, {Frayer}, {Christensen},
  {Gallazzi}, {Michalowski}, {Richard}, \& {Zavala}}]{Knudsen16}
{Knudsen}, K.~K., {Watson}, D., {Frayer}, D., {Christensen}, L., {Gallazzi},
  A., {Michalowski}, M.~J., {Richard}, J., \& {Zavala}, J. 2016, ArXiv e-prints

\bibitem[{{Larson}(1981)}]{Larson81}
{Larson}, R.~B. 1981, \mnras, 194, 809

\bibitem[{{Lippok} {et~al.}(2016){Lippok}, {Launhardt}, {Henning}, {Beuther},
  {Kainulainen}, {Krause}, {Linz}, {Nielbock}, {Ragan}, {Robitaille},
  {Sadavoy}, \& {Schmiedeke}}]{Lippok16}
{Lippok}, N., {et~al.} 2016, ArXiv e-prints

\bibitem[{{Maiolino} {et~al.}(2015){Maiolino}, {Carniani}, {Fontana},
  {Vallini}, {Pentericci}, {Ferrara}, {Vanzella}, {Grazian}, {Gallerani},
  {Castellano}, {Cristiani}, {Brammer}, {Santini}, {Wagg}, \&
  {Williams}}]{Maiolino15}
{Maiolino}, R., {et~al.} 2015, \mnras, 452, 54

\bibitem[{{Meurer} {et~al.}(1999){Meurer}, {Heckman}, \& {Calzetti}}]{Meurer99}
{Meurer}, G.~R., {Heckman}, T.~M., \& {Calzetti}, D. 1999, \apj, 521, 64

\bibitem[{{Meurer} {et~al.}(1995){Meurer}, {Heckman}, {Leitherer}, {Kinney},
  {Robert}, \& {Garnett}}]{Meurer95}
{Meurer}, G.~R., {Heckman}, T.~M., {Leitherer}, C., {Kinney}, A., {Robert}, C.,
  \& {Garnett}, D.~R. 1995, \aj, 110, 2665

\bibitem[{{Micha{\l}owski} {et~al.}(2010){Micha{\l}owski}, {Murphy}, {Hjorth},
  {Watson}, {Gall}, \& {Dunlop}}]{Michalowski10}
{Micha{\l}owski}, M.~J., {Murphy}, E.~J., {Hjorth}, J., {Watson}, D., {Gall},
  C., \& {Dunlop}, J.~S. 2010, \aap, 522, A15

\bibitem[{{Nozawa} {et~al.}(2007){Nozawa}, {Kozasa}, {Habe}, {Dwek}, {Umeda},
  {Tominaga}, {Maeda}, \& {Nomoto}}]{Nozawa07}
{Nozawa}, T., {Kozasa}, T., {Habe}, A., {Dwek}, E., {Umeda}, H., {Tominaga},
  N., {Maeda}, K., \& {Nomoto}, K. 2007, \apj, 666, 955

\bibitem[{{Ouchi} {et~al.}(1999){Ouchi}, {Yamada}, {Kawai}, \&
  {Ohta}}]{Ouchi99}
{Ouchi}, M., {Yamada}, T., {Kawai}, H., \& {Ohta}, K. 1999, \apjl, 517, L19

\bibitem[{{Pallottini} {et~al.}(2014){Pallottini}, {Ferrara}, {Gallerani},
  {Salvadori}, \& {D'Odorico}}]{Pallottini14}
{Pallottini}, A., {Ferrara}, A., {Gallerani}, S., {Salvadori}, S., \&
  {D'Odorico}, V. 2014, \mnras, 440, 2498

\bibitem[{{Planck Collaboration} {et~al.}(2015){Planck Collaboration}, {Ade},
  {Aghanim}, {Arnaud}, {Ashdown}, {Aumont}, {Baccigalupi}, {Banday},
  {Barreiro}, {Bartlett}, \& et~al.}]{Ade15}
{Planck Collaboration} {et~al.} 2015, ArXiv e-prints

\bibitem[{{Reddy} {et~al.}(2012){Reddy}, {Dickinson}, {Elbaz}, {Morrison},
  {Giavalisco}, {Ivison}, {Papovich}, {Scott}, {Buat}, {Burgarella},
  {Charmandaris}, {Daddi}, {Magdis}, {Murphy}, {Altieri}, {Aussel},
  {Dannerbauer}, {Dasyra}, {Hwang}, {Kartaltepe}, {Leiton}, {Magnelli}, \&
  {Popesso}}]{Reddy12}
{Reddy}, N., {et~al.} 2012, \apj, 744, 154

\bibitem[{{Riechers} {et~al.}(2014){Riechers}, {Carilli}, {Capak}, {Scoville},
  {Smol{\v c}i{\'c}}, {Schinnerer}, {Yun}, {Cox}, {Bertoldi}, {Karim}, \&
  {Yan}}]{Riechers14}
{Riechers}, D.~A., {et~al.} 2014, \apj, 796, 84

\bibitem[{{Schneider} {et~al.}(2016){Schneider}, {Bontemps}, {Motte},
  {Ossenkopf}, {Klessen}, {Simon}, {Fechtenbaum}, {Herpin}, {Tremblin},
  {Csengeri}, {Myers}, {Hill}, {Cunningham}, \& {Federrath}}]{Schneider15}
{Schneider}, N., {et~al.} 2016, \aap, 587, A74

\bibitem[{{Stutz} {et~al.}(2010){Stutz}, {Launhardt}, {Linz}, {Krause},
  {Henning}, {Kainulainen}, {Nielbock}, {Steinacker}, \& {Andr{\'e}}}]{Stutz10}
{Stutz}, A., {et~al.} 2010, \aap, 518, L87

\bibitem[{{Takeuchi} {et~al.}(2012){Takeuchi}, {Yuan}, {Ikeyama}, {Murata}, \&
  {Inoue}}]{Takeuchi12}
{Takeuchi}, T.~T., {Yuan}, F.-T., {Ikeyama}, A., {Murata}, K.~L., \& {Inoue},
  A.~K. 2012, \apj, 755, 144

\bibitem[{{Todini} \& {Ferrara}(2001)}]{Todini01}
{Todini}, P., \& {Ferrara}, A. 2001, \mnras, 325, 726

\bibitem[{{Vallini} {et~al.}(2016){Vallini}, {Ferrara}, {Pallottini}, \&
  {Gallerani}}]{Vallini16}
{Vallini}, L., {Ferrara}, A., {Pallottini}, A., \& {Gallerani}, S. 2016, ArXiv
  e-prints

\bibitem[{{Wagg} {et~al.}(2009){Wagg}, {Kanekar}, \& {Carilli}}]{Wagg09}
{Wagg}, J., {Kanekar}, N., \& {Carilli}, C.~L. 2009, \apjl, 697, L33

\bibitem[{{Watson} {et~al.}(2015){Watson}, {Christensen}, {Knudsen}, {Richard},
  {Gallazzi}, \& {Micha{\l}owski}}]{Watson15}
{Watson}, D., {Christensen}, L., {Knudsen}, K.~K., {Richard}, J., {Gallazzi},
  A., \& {Micha{\l}owski}, M.~J. 2015, \nat, 519, 327

\bibitem[{{Weingartner} \& {Draine}(2001)}]{Weingartner01}
{Weingartner}, J.~C., \& {Draine}, B.~T. 2001, \apj, 548, 296

\end{thebibliography}

\newpage 
\label{lastpage} 
\end{document}